\documentclass[aps,showpacs,preprintnumbers,showkeys,twocolumn]{revtex4-1}%
\usepackage{amsfonts,amscd}
\usepackage{amsmath}
\usepackage{amssymb}
\usepackage{graphicx}
\usepackage{amsfonts}%
\providecommand{\U}[1]{\protect\rule{.1in}{.1in}}
\newtheorem{theorem}{Theorem}

\newtheorem{claim}[theorem]{Claim}

\newtheorem{criterion}[theorem]{Criterion}

 \DeclareMathOperator{\tr}{Tr}
 
\begin{document}
\title{Undetermined states: how to find them and their applications}
\author{Min-Hsiu Hsieh,$^{1}$ Wen-Tai Yen,$^{2}$ and Li-Yi Hsu$^{2}$}
\affiliation{$^{1}$ERATO-SORST Quantum Computation and Information Project, Japan Science
and Technology Agency 5-28-3, Hongo, Bunkyo-ku, Tokyo, Japan}
\affiliation{$^{2}$Department of physics, Chung Yuan Christian University, Chungli, Taiwan}

\begin{abstract}
We investigate the undetermined sets consisting of two-level, multi-partite
pure quantum states, whose reduced density matrices give absolutely no
information of their original states.\ Two approached of finding these quantum
states are proposed. One is to establish the relation between codewords of the
stabilizer quantum error correction codes (SQECCs) and the undetermined
states. The other is to study the local complementation rules of the graph
states. As an application, the undetermined states can be exploited in the
quantum secret sharing scheme. The security is gauranted by their undeterminedness.

\end{abstract}
\volumeyear{year}
\volumenumber{number}
\issuenumber{number}
\eid{identifier}
\date[Date text]{date}
\received[Received text]{date}

\revised[Revised text]{date}

\accepted[Accepted text]{date}

\published[Published text]{date}

\maketitle

\section{Introduction}

Quantum entanglement has been regarded as an useful physical resource in
quantum information and quantum computation. Different non-local correlations
embedded in entanglement are employed for different tasks. Characterization of
quantum entanglement plays a fundamental role in quantum information science.
Entanglement measures in various aspects of research purposes have been
proposed \cite{1}. Therein, an interesting and natural approach, as originally
proposed by Linden \emph{et. al.}, is to quantify information contained in the
reduced density matrix of a given state \cite{2}. In other words, it is
concerned whether the partial or low-ordered correlations of the reduced
matrices can fully reveal the high-ordered correlations. If the answer is
positive, these entangled states, either pure or mixed, can be completely
determined by their parts. Surprisingly, \emph{almost }every pure three-qubit
entangled state can be entirely determined by its two-qubit reduced density
matrices \cite{2}. Later the solution of constructing the three-qubit pure
state using any two two-qubit reduced density matrices was explicitly
illustrated \cite{22}. Furthermore, Linden and Wootters showed that a fraction
of quantum particles of \emph{almost} $n$-party multi-level pure entangled
states can reveal as much information as that of all qubits \cite{21}.

In other words, only a few pure entangled states whose parts cannot determine
the whole. As a trivial example, the four orthogonal two-qubit Bell states are
undetermined since their one-qubit reduced density matrices are all maximally
mixed. Recently, Walck and Lyons rigorously proved that the generalized
$n$-qubit Greenberger-Horne-Zeilinger (GHZ) states and their local unitary
equivalents are the only states that are not uniquely determined among pure
states by their ($n-1$)-qubit reduced density matrices \cite{4,5}. In this
paper, we focus on finding undetermined sets of pure quantum states.
Specifically, we would like to construct an undetermined set of $n$-qubit
quantum stabilizer\ states such that after tracing out arbitrary $D$ qubits,
the resulting reduced density matrices are all the same. In quantum
information science, characterizing these undetermined sets is more important
since it strongly relates to applications in quantum cryptography.

In this paper, we propose two approaches of finding the undetermined
stabilized states. One is to exploit the logic states of stabilizer quantum
error-correcting codes (SQECCs) to comprise the undetermined sets. The other
approach is based on the local equivalence of graph states. Our paper is
organized as follows. First, some definitions are given in Sec. II. The
properties of SQECCs are briefly reviewed. Specifically, we focus on
$[[n,1,d]]$ SQECCs. The undetermineness of\ codeword states of the SQECCs are
studied. Inspired by $[[n,1,d]]$ SQECCs, some other stabilizer states as
undetermined states can be also found. Next, in Sec. III, graph states are
introduced. Therein, the graph transformation, called local complementation
helps find the conditionally undetermined sets of graph states. Finally, in
Sec. IV, we demonstrate several applications of these undetermined states in
quantum cryptography. In the following, $X_{i}$, $Y_{i}$, and $Z_{i}$ denote
the Pauli operators $\sigma_{x}$, $\sigma_{y}$ and $\sigma_{z}$ on the $i$-th
qubit, respectively.

\section{Unconditionally undetermined sets of quantum states}

\label{III}

Throughout this paper, an undetermined set ($D_{u}$-set hereafter) of two or
more $n$-qubit pure states is called unconditionally $D$-undetermined, if the
reduced density matrices of \emph{arbitrary} ($n-D$) qubits are always the
same for all state in the $D_{u}$-set. According to the definition, the
following claim comes up immediately.

\begin{claim}
If a $D_{u}$-set consists of two pure states, $|\psi_{0}\rangle$ and
$|\psi_{1}\rangle$, then for arbitrary $D$-qubit set $(i_{1},i_{2}%
,\cdots,i_{D})$, there is a unitary $U_{i_{1}i_{2}\cdots i_{D}}$ such that
$|\psi_{1}\rangle=U_{i_{1}i_{2}\cdots i_{D}}|\psi_{0}\rangle$.
\end{claim}

The proof is straightforward and readers can refer to Ref. \cite{NC,4}. In
general, for two arbitrary $D$-qubit sets ($i_{1}$, $i_{2}$, $\cdots$, $i_{D}%
$) and ($i_{1}^{\prime},i_{2}^{\prime},\cdots,i_{D}^{\prime}$), there must
exist some non-trivial $D$-qubit unitary matrices $U_{i_{1}i_{2}\cdots i_{D}}$
and $U_{i_{1}^{\prime}i_{2}^{\prime}\cdots i_{D}^{\prime}}$, such that
\begin{equation}
|\psi_{1}\rangle=U_{i_{1}i_{2}\cdots i_{D}}|\psi_{0}\rangle\text{ and }%
|\psi_{1}\rangle=U_{i_{1}^{\prime}i_{2}^{\prime}\cdots i_{D}^{\prime}}%
|\psi_{0}\rangle. \label{U}%
\end{equation}
Obviously,
\begin{equation}
U_{i_{1}i_{2}\cdots i_{D}}^{-1}U_{i_{1}^{\prime}i_{2}^{\prime}\cdots
i_{D}^{\prime}}\left\vert \psi_{0}\right\rangle =\left\vert \psi
_{0}\right\rangle .
\end{equation}
Notably, here non-trivial $U_{i_{1}i_{2}\cdots i_{D}}$ and $U_{i_{1}^{\prime
}i_{2}^{\prime}\cdots i_{D}^{\prime}}$ cannot be decomposed as tensor product
concluding any 1-qubit identity. Seemingly, $\left\vert \psi_{0}\right\rangle
$ is stabilized by $U_{i_{1}i_{2}\cdots i_{D}}^{-1}U_{i_{1}^{\prime}%
i_{2}^{\prime}\cdots i_{D}^{\prime}}$ and hence the claim implicitly sheds an
insight of finding undetermined states via stabilizer states

\subsection{Review of SQECCs}

Before further proceeding, we introduce SQECCs as follows. A quantum code that
encodes $k$ logical qubits into $n$ physical qubits and has distance $d$ is
denoted as $[[n,k,d]]$. For an $[[n,1,$ $d]]$ stabilizer quantum
error-correcting code ($k=1$), the logic states or, equivalently, codewords
\ are denoted by $\left\vert \overline{0}\right\rangle $ and $\left\vert
\overline{1}\right\rangle $, respectively. The code space $\mathcal{C}$,
spanned by $\{|\overline{0}\rangle,|\overline{1}\rangle\}$, was fixed by its
stabilizer group $\mathcal{S}=\{g_{1},g_{2},\cdots,g_{n-1}\}$, where
$g_{1},g_{2},\cdots,g_{n-1}$ are generators of the group \cite{NC}. Moreover,
$g_{i}$, $i=1,2,\cdots,n-1$, are non-trivial n-fold Pauli operators in
$\mathcal{P}_{n}=\langle i,Z_{1},X_{1},\cdots,Z_{n},X_{n}\rangle$ and they all
commute with each other. Then $\forall M\in\mathcal{S}$, and $\forall
|\psi\rangle\in\mathcal{C}$, we have
\[
M|\psi\rangle=|\psi\rangle.
\]
Therefore, the codeword state $|\psi\rangle$ is also called the
\textquotedblleft stabilizer state\textquotedblright.

The normalizer of $\mathcal{S}$, denoted as $\mathcal{N}(\mathcal{S})$, is the
subgroup of $\mathcal{P}_{n}$ that commutes with every element of
$\mathcal{S}$. In the case of an $[[n,1,d]]$ SQECC, we denote its normalizer
as $\mathcal{N}(\mathcal{S})=\langle i,\overline{Z},\overline{X}\rangle$,
where $\overline{Z}$ and $\overline{X}$ are called the logic phase-flip
operator and the logic bit-flip operator, respectively \cite{DG97thesis}:
\begin{align*}
\overline{Z}|\overline{0}\rangle &  =|\overline{0}\rangle\\
\overline{Z}|\overline{1}\rangle &  =-|\overline{1}\rangle\\
\overline{X}|\overline{0}\rangle &  =|\overline{1}\rangle\\
\overline{X}|\overline{1}\rangle &  =|\overline{0}\rangle.
\end{align*}

Notice that there is some degree of freedom for choosing the logic bit-flip
operator, denoted by $\overline{X}$. Let the operator $X$ satisfy the
condition
\begin{equation}
X\in\mathcal{N}/\mathcal{S}\text{ and }\{X\text{, }\overline{Z}\}=0,
\end{equation}
where $\{$ , $\}$ is anti-commutator. We can also choose any of $X\mathcal{S}$
or $X$ as $\overline{X}$. Here we define the set of all legitimate
$\overline{X}$ operators as $\mathcal{X},$ where
\begin{equation}
\mathcal{X}=\{X|X\in\mathcal{N}/\mathcal{S}\text{ and }\{X,\overline{Z}\}=0\}.
\end{equation}

By definition, given an $[[n,1,d]]$ SQECC, there must exist \emph{at least}
one $d$-qubit undetected error
\begin{equation}
\mathcal{E}_{i_{1}i_{2}\cdots i_{d}}=(\mathcal{E}_{i_{1}}\otimes
\mathcal{E}_{i_{2}}\otimes\cdots\otimes\mathcal{E}_{i_{d}}), \label{LOO}%
\end{equation}
where $\mathcal{E}_{i}$ is a non-identity Pauli operator on qubit $i$, such
that
\begin{equation}
\mathcal{E}_{i_{1}i_{2}\cdots i_{d}}\left\vert \overline{0}\right\rangle
=\left\vert \overline{1}\right\rangle .
\end{equation}
Obviously, $\mathcal{E}_{i_{1}i_{2}\cdots i_{d}}\in\mathcal{X}$.

\subsection{Logic states as elements of the $D_{u}$-set}

Before presenting new sets of undetermined states, it should be noted that the
unitary $U_{i_{1}i_{2}\cdots i_{D}}$ in Eq. (\ref{U}) is not necessarily as
tensor product of $D$ non-identity Pauli operators. In the following, in order
to apply the theory of stabilizer quantum error correction, we restrict
$U_{i_{1}i_{2}\cdots i_{D}}$ as the format of $D$-fold Pauli operators. Now,
the connection between $[[n,1,d]]$ SQECC and $D_{u}$-set can be described as
follows. Two pure states \ $|\psi_{0}\rangle$ and $|\psi_{1}\rangle$ in the
above claim can be presumed as $\left\vert \overline{0}\right\rangle $ and
$\left\vert \overline{1}\right\rangle $ of the $[[n,1,d]]$ SQECC. In addition,
the unitary operators $U_{i_{1}i_{2}\cdots i_{d}}$ and $U_{i_{1}^{\prime}%
i_{2}^{\prime}\cdots i_{d}^{\prime}}$ in Eq. (\ref{U}) each is equal to
$\mathcal{E}_{i_{1}i_{2}\cdots i_{d}}$ in Eq. (\ref{LOO}) as sthe logic
bit-flip operator. In order for logic states comprising the $D_{u}$-set ,
however, the following criterion must be fulfilled according to the claim.

\begin{criterion}
For arbitray $D$ qubits, there must exist an an undetected error operator
$\mathcal{E}_{i_{1}i_{2}\cdots i_{D}}\in\mathcal{X}$ as $U_{i_{1}i_{2}\cdots
i_{d}}.$
\end{criterion}

We emphasize that \emph{not} two logic states of any $[[n,1,d]]$ SQECCs can
always comprise a two-element $D_{u}$-set. As a result, we propose how to find
a two-element $D_{u}$-set as follows. (i) choose an $[[n,1,d]]$ SQECC. (ii)
verify whether this SQECC satisfies the criterion. If the criterion is
satisfied, the logic states can comprise a $D_{u}$-set.\ 

As a trivial example, here we show how to recover the result of \cite{4} using
the idea of SQECCs. Wherein, the $n$-qubit GHZ states $\frac{1}{\sqrt{2}%
}(\left\vert 0\right\rangle ^{\otimes n}+\left\vert 1\right\rangle ^{\otimes
n})$ and $\frac{1}{\sqrt{2}}(\left\vert 0\right\rangle ^{\otimes n}-\left\vert
1\right\rangle ^{\otimes n})$ can be regarded as the logic states of the
$[[n,1,1]]$ SQECC, $\left\vert \overline{0}\right\rangle $ and $\left\vert
\overline{1}\right\rangle $, respectively. The corresponding stabilizer
generators $g_{i}$ is $Z_{i}Z_{i+1}$, $i\in\{1,2,\cdots,n-1\}$, and the logic
phase-flip operator $\overline{Z}$ can be chosen as $X^{\otimes n}.$ It is
noteworthy that such an $[[n,1,1]]$ stabilizer code can correct all the
single-qubit bit-flip errors, but not single-qubit phase-flip errors.
Obviously, $Z_{i}\in\mathcal{N}$ and $\{Z_{i}$, $\overline{Z}\}=0$. In other
words, there exist Pauli operators $Z_{i}\in$ $\mathcal{X}$, $\forall i$, on
each qubit such that
\begin{equation}
|\overline{1}\rangle=Z_{i}|\overline{0}\rangle\text{ and }|\overline{0}%
\rangle=Z_{i}|\overline{1}\rangle.\text{ }%
\end{equation}
Therefore, these GHZ states $|\overline{0}\rangle$ and $|\overline{1}\rangle$
are 1-undetermined states according to the claim.

In addition, let $E_{D}$ be the number of operators in $\mathcal{X}$ whose
weight is $D$. The inequality,
\[
E_{D}\geq\frac{n!}{D!(n-D)!},
\]
is a necessary condition for an $[[n,1,d]]$ SQECC such that its codewords can
comprise $D$-set. It is also easy to see that
\[
D\geq d.
\]
In the case of $n$-qubit GHZ states, $D=d=1.$ In the following, some logic
states as elements of a $D_{u}$-set are given.

\subsection{ $[[4,1,2]]$ SQECC}

As a simple example, firstly, we consider the $[[4,1,2]]$ SQECC with the
following generators
\[
g_{1}=Y_{1}Y_{3},\text{ }g_{2}=Y_{2}Y_{4},\text{ and }g_{3}=Z_{1}Z_{2}%
Z_{3}Z_{4}.
\]
Here we pick $\overline{Z}=Y_{1}X_{2}Z_{4}$. Then it is easy to verify that
\[
\{Y_{1}Y_{2}\text{, }Y_{2}Y_{3}\text{, }Y_{3}Y_{4}\text{, }Y_{4}Y_{1}\text{,
}Z_{1}Z_{3}\text{, }X_{2}X_{4}\}\in\mathcal{X}.
\]
As a result, the two-qubit reduced density of the corresponding logic states
$\left\vert \psi_{0}\right\rangle $ and $\left\vert \psi_{1}\right\rangle $
are always the same. In other words, $\left\vert \psi_{0}\right\rangle $ and
$\left\vert \psi_{1}\right\rangle $ are 2-undetermined.

In general, there exist several $[[4,1,2]]$ SQECCs that are not equivalent.
The set of generators is constructed based on a computer search with the
constrain that $\mathcal{X}$ must contains all possible two-qubit Pauli operators.

\subsection{ $[[5,1,3]]$ cyclic SQECC and its generalization}

Another set comprises the logic states of the $[[5,1,3]]$ SQECC \cite{513},
where the stabilizer generator
\[
g_{i}=\mathfrak{R}^{i-1}(X_{1}Z_{2}Z_{3}X_{4}I_{5})
\]
($\mathfrak{R}$ is the right-shift operator), and
\[
\overline{Z}=Z_{1}Z_{2}Z_{3}Z_{4}Z_{5},
\]
respectively. Interestingly, we have the error set
\[
\{Y_{2}Y_{3}X_{5}\text{+ cyclic terms, }X_{1}Z_{2}Z_{5}\text{+ cyclic terms
}\}\in\mathcal{X}.
\]
Therefore, the logic states are 3-undetermined states.

Inspired by the $[[5,1,3]]$ code, we consider the following $n$-qubit states
$\left\vert \psi_{0}\right\rangle $ and $\left\vert \psi_{1}\right\rangle $.
Wherein, the stabilizer group is generated by $g_{i}$, where $g_{i}%
=\mathfrak{R}^{i-1}(X^{\otimes p}Z^{\otimes2p}X^{\otimes p}I^{\otimes q})$ and
the logic phase-flip operator $\overline{Z}$ is $Z^{\otimes(4p+q)}$,
respectively. Note that $n=4p+q$, where $q\equiv n$ $\operatorname{mod}$ 4.
For $5\leq n\leq15$, it is found that when arbitrary $n-2$ qubits are traced
out, the reduced density matrices are the same and undetermined.

\section{Conditionally undetermined sets of quantum states}

We can weaken the unconditionally undetermined condition as follows. The
$D_{c}^{\prime}$- set consisting of pure quantum states is called
conditionally undetermined, if, when some particular $D^{\prime}$ qubits are
traced out, the reduced density matrices are the same. This particular set
corresponds to the \emph{unauthorized set} first mentioned in \cite{Gottesman}%
. Obviously, the logic states of any $[[n,1,d]]$ SQECC comprise a
$D_{c}^{\prime}$- set with the minimal $D^{\prime}$ equal to $d$. In addition,
if the $D_{c}^{\prime}$- set is identical to some $D_{u}$- set, we have $d\leq
D^{\prime}\leq D$. As a concrete example, we consider the $[[7,1,3]]$ Steane
code \cite{713}, where the six generators are
\begin{multline}
X_{3}X_{4}X_{5}X_{6},X_{2}X_{3}X_{6}X_{7},X_{1}X_{3}X_{5}X_{7},\\
Z_{3}Z_{4}Z_{5}Z_{6},Z_{2}Z_{3}Z_{6}Z_{7},Z_{1}Z_{3}Z_{5}Z_{7},
\end{multline}
and $\overline{Z}=Z_{1}Z_{2}Z_{3}Z_{4}Z_{5}Z_{6}Z_{7}$. It is easy to see that
$|\psi_{0}\rangle$ and $|\psi_{1}\rangle$ are conditionally $3$-undetermined
states since $\tr_{(234)}(|\psi_{0}\rangle\!\langle\psi_{0}|)\neq
\tr_{(234)}(|\psi_{1}\rangle\!\langle\psi_{1}|)$. By further check, these two
logic states become unconditionally $5$-undetermined if arbitrary five qubits
are traced out.

\subsection{Review of graph states}

In addition, another approach of finding $D_{c}^{\prime}$- sets is based on
the equivalent graph states under local transformation. Before further
proceeding we review graph states as follows. In general, an $n$-vertex graph
is denoted by $\mathrm{G}_{n}=(\mathrm{V},\mathrm{E})$, which can be composed
of a set $\mathrm{V}$ of $n$ vertices and a set $\mathrm{E}$ of edges.The
neighboring set of the vertex $a$, $\mathrm{N}(a)$, is defined as
$\mathrm{N}(a)=\{i|$ ($a$, $i$) $\in\mathrm{E}\}$. The quantum state
corresponding to $\mathrm{G}_{n}$ is denoted by $\left\vert \mathrm{G}%
_{n}\right\rangle $. Therein, the vertex $i$ correpsonds to the qubit $i$,
$\forall$ $i\in\{1,\ldots,n\}$. In addition, $\left\vert \mathrm{G}%
_{n}\right\rangle $ is stabilized by the stabilizer generators, $\mathrm{g}%
_{i}$
\begin{equation}
\mathrm{g}_{i}=X_{i}\prod_{j\in N(i)}Z_{j},\forall i=1,\cdots,n,\label{3}%
\end{equation}

As an example, consider the four-qubit graph state $\left\vert \mathrm{G}%
_{4}\right\rangle $ with associated graph $\mathrm{G}_{4}=(\mathrm{V}%
,\mathrm{E})$. Wherein, there are four vertices 1, 2, 3 and 4, and
$\mathrm{E}=\{$(1, 2), (2, 3), (3, 4), (4, 1), (2,4)$\,\}$. Another genuine
four-qubit entangled state $\left\vert \chi\right\rangle $ has the form
\[
\left\vert \chi\right\rangle =\frac{1}{\sqrt{2}}(\left\vert \phi
^{0}\right\rangle +\left\vert \phi^{1}\right\rangle )
\]
with
\begin{equation}
\left\vert \phi^{0}\right\rangle =\frac{1}{2}(\left\vert 0000\right\rangle
-\left\vert 0011\right\rangle -\left\vert 0101\right\rangle +\left\vert
0110\right\rangle )
\end{equation}%
\begin{equation}
\left\vert \phi^{1}\right\rangle =\frac{1}{2}(\left\vert 1001\right\rangle
+\left\vert 1010\right\rangle +\left\vert 1100\right\rangle +\left\vert
1111\right\rangle )
\end{equation}
It can be easily checked that $\tr_{24}(|\chi\rangle\!\langle\chi
|)=\tr_{24}(|\mathrm{G}_{4}\rangle\!\langle\mathrm{G}_{4}|)$ \cite{pla}.

Local complementation $\mathcal{L(}a\mathcal{)}$ on $\mathrm{G}_{n}%
=(\mathrm{V},\mathrm{E})$ is to complement the connection relation in the
vertex set $\mathrm{N}(a)$. In details, for any two vertices $i$, $j\in$
$\mathrm{N}(a)$, $\mathcal{L(}a\mathcal{)}$ removes (creates)\ the edge ($i$,
$j$) if ($i$, $j$) $\in\mathrm{E}$ ($\notin\mathrm{E}$) \cite{0602096}.
Correspondingly, two graph states are local equivalent, if these two
associated graphs are equivalent under the sequential local complementations.
If two graphs $\mathrm{G}_{n}=(\mathrm{V},\mathrm{E})$ and $\mathrm{G}%
_{n}^{\prime}=(\mathrm{V}^{\prime},\mathrm{E}^{\prime})$ are local equivalent
under the operation $\mathcal{L(}a\mathcal{)}$, we have%

\begin{equation}
\left\vert \mathrm{G}_{n}\right\rangle =U_{\mathrm{N}(a)\cup a}\left\vert
\mathrm{G}_{n}^{\prime}\right\rangle ,
\end{equation}
and as a result,%

\begin{equation}
\tr_{N(a)\cup a}(|\mathrm{G}_{n}\rangle\!\langle\mathrm{G}_{n}|)=\tr_{N(a)\cup
a}(|\mathrm{G}_{n}\rangle\!\langle\mathrm{G}_{n}^{\prime}|)\text{. }%
\end{equation}
A minimal $D_{c}^{\prime}$-set has the form
\[
D_{c}^{\prime}=\{\left\vert \mathrm{G}_{n}\right\rangle ,\left\vert
\mathrm{G}_{n}^{\prime}\right\rangle \}\text{ with }D^{\prime}=\left\vert
\mathrm{N}(a)\right\vert +1.
\]

\section{Applications of undetermined states in quantum cryptography}

In the following, we will show that the undetermined states have several
important applications in quantum cryptography.

\subsection{Quantum bit commitment}

A well-known result is\ that the unconditionally secure quantum bit commitment
(QBC) is impossible \cite{qbc}. In brief, to defy QBC, the dishonest sender
initially prepares the singlet state, $|B\rangle$. In the sealing phase, the
sender keeps one qubit and seals the other qubit into the black box, which is
then delivered to the honest receiver.

In the detailed analysis by Lo and Chau \cite{qbc}, the states $|B\rangle$ and
$U\otimes I\left\vert B\right\rangle $ are undetermined by their single-qubit
density matrices, where $U$ \ and $\ I$ are the single qubit operation and
$2\times2$ identity matrix, respectively. That is, the dishonest sender can
perform the single qubit operation $U$ on the qubit at hand without the
receiver's awareness. On the other hand, the sender can access the full
information via the anti-perfect correlation of the singlet state. As a
result, the undetermineness of two-qubit\ Bell states rejects the security of QBC.

\subsection{Quantum secret sharing (QSS)}

In brief, classical\ secret sharing scheme (SSS) refers to a method of
splitting secret meassage into $n$ shares, each containing only partial
information. The QSS protocol for sharing classical secrets is first proposed
by Hillery, Bu\v{z}ek, and Berthiaume (HBB) \cite{3}, which can be briefed as
follows. (i) The distribution phase: the sender Alice \emph{always} initially
prepares the same three-qubit GHZ states, $\frac{1}{\sqrt{2}}(\left\vert
0\right\rangle ^{\otimes3}+\left\vert 1\right\rangle ^{\otimes3})$ in each
round. Then she sends two of these three qubits to each the receivers, Bob and
Charlie. Afterwards, all three parties perform either $Y$ or $X$ measurement
at random. The receivers each send the measurement basis to Alice. On the
other hand, Alice performs her measurement. Here the secret bit is Alice's
measurement outcome. (ii) The revealing phase: Alice announces her measurement
basis. If either one or three $X$ measurements are preformed among these three
parties, they keep their respective outcomes, otherwise these outcomes are
discarded. As for the reserved outcomes, Bob and Charlie have to collaborate
to deduce Alice' outcome. To achieve this, they just reveal their respective
outcomes to each other in private. At last, in order to detect the
eavesdropping in the checking phase, some portion of the outcomes should be
chosen randomly and then announced public. If the bit error is beyond some
threshold, there must exist eavesdropping and hence the secret bits should be abandoned.

Recently, cryptanalysis of the HBB protocol using the intercept-and-resend
attack is proposed by Qin \emph{et. al.} \cite{Ins}. Without loss of
generality, the dishonest Charlie is presumed equipped with unlimited power
without violating the physics principle. To access the secret bit without
Bob's assistance, Charlie intercepts both the delivered qubits and performs
joint operations on the these qubits with unlimited ancilla qubits. He then
resends one qubit to Bob. In the revealing phase, the dishonest Charlie can
forge his measurement basis before the actual measurement. In addition,
Charlie can wiretap Bob's measurement basis. Finally, after Alice's
announcement, Charlie himself can deduce Alice's outcome without any
assistance and awareness \cite{Ins}.

Here we demonstrate that the undetermined states can save the security of the
HBB protocol with any other further modification. The essential weakness of
the original protocol lies in the changeless state prepared in the
distribution phase. To rescue its security, the QSS protocol can be modified
as follows. In the preparation phase of each round, one of the logic states of
$[[3,1,1]]$ stabilizer code, which are $|\psi_{0}\rangle=\frac{1}{\sqrt{2}%
}(|0\rangle^{\otimes3}+|1\rangle^{\otimes3})$ and $|\psi_{1}\rangle=\frac
{1}{\sqrt{2}}(|0\rangle^{\otimes3}-|1\rangle^{\otimes3})$, are initially
prepared with equal probability. Then two qubits of the prepared three-qubit
state are sent to the receivers. In the revealing phase, Alice announces which
state is prepared after she ensures that Bob and Charlie have exchanged their
measurement outcomes. The proof of security is analogue to the impossibility
of the secure quantum bit commitment as addressed above \cite{qbc}.
Essentially, the states $|\psi_{0}\rangle$ and $|\psi_{1}\rangle$, which are
randomly distributed, are 1-undetermined. It is impossible for Bob and Charlie
to discriminate $|\psi_{0}\rangle$ and $|\psi_{1}\rangle$ from their two-qubit
density matrices. That is, the undeterminedness as the physics principle
prevents a successful eavesdropping. In this condition, any forge can be
regarded as random guess, which can be detected at the checking phase with
half probability. Therefore, the modified quantum secret sharing scheme is
unconditionally secure.

Furthermore, the HBB protocol be generalized as follows. (i) In the
distribution phase, the dealer can prepare one of the $n$-qubit states in
the$\ D_{u}$-set and then distrbutes $n-D$ qubits each into $n-D$ receivers.
(ii) In the revealing phase, $n-D$ of $n$\ receivers discuss their outcomes of
the local measurements before sender's announcement on state preparation and
the local measurement.

The security of the above QSS\ protocols is gauranteed by the undetermineness
of the elements in the$\ D_{u}$-set. As a result, even intercepting all
delivered qubits, any dishonest receiver cannot determine which states are
distributed. Using the protocol similar to the above modified protocol, the
correlation embedded in quantum entanglement can exploited for secret sharing.
For example, as pointed out, for any graph-state-based QSS scheme for
classical bits \cite{cc} cannot be secure, if only one element of the $D_{u}%
$-set or $D_{c}^{\prime}$-set is determinstically prepared \cite{cjp}.

Finally, the $D$-undetermined states provide unconditional security for a
quantum $((n-D+1,n))$ threshold scheme if such scheme exists \cite{Gottesman}.
A quantum $((k,n))$ threshold scheme has $n$ shares, of which \emph{any} $k$
shares are sufficient to reconstruct the secret, while any set of $k-1$ or
fewer shares has no information about the secret. This can be easily verified
because that the reduced density matrix of any set of parties less than or
equal to $D$ is the same. Furthermore, the conditionally $D^{\prime}%
$-undetermined states from some SQECCs provide us a trivial way of determining
the unauthorized sets (therefore its quantum access structure \cite{Gottesman}%
) when we use the codewords of this SQECC in the quantum secret sharing protocol.

\section{Conclusion}

In this paper, we focus on the construction of families of undetermined
\textit{pure} state. However, the undetermined mixed states can be also
constructed from SQECCs similarly. Consider the $[[n,2,d]]$ SQECCs with the
logic states $\left\vert \overline{ij}\right\rangle $, $i,j\in\{0,1\}$. Let
$\overline{Z}_{1}$ and $\overline{Z}_{2}$ be the logic $\sigma_{z}$ operators
on logic states $|\bar{\imath}\rangle$ and $|\bar{j}\rangle$, respectively. In
addition, denote the set
\[
\mathcal{X}_{i}=\{X_{i}|X_{i}\in\mathcal{N}\ \text{and}\ \{X_{i},\overline
{Z}_{i}\}=0\},
\]
where $i\in\{1,2\}$, and $\mathcal{N}$ is the normalizer of the stabilizer
group generated by $g_{1},$ $g_{2},\cdots$, $g_{n-2}$. Now, define the density
matrices of two mixed states
\[
\rho_{0}=\frac{1}{2}(\left\vert \overline{00}\right\rangle \left\langle
\overline{00}\right\vert +\left\vert \overline{11}\right\rangle \left\langle
\overline{11}\right\vert ),
\]
and
\[
\rho_{1}=\frac{1}{2}(\left\vert \overline{10}\right\rangle \left\langle
\overline{10}\right\vert +\left\vert \overline{01}\right\rangle \left\langle
\overline{01}\right\vert ).
\]
If the mixed states $\rho_{0}$ and $\rho_{1}$ are $D$-undetermined, by
definition, for arbitrary $D$ qubits $i_{1}$, $i_{2}$, $\cdots$, $i_{D}$,
there must be an undetected error $\mathcal{E}_{i_{1}i_{2}\cdots i_{D}}$, such
that
\[
\mathcal{E}_{i_{1}i_{2}\cdots i_{D}}\in\mathcal{X}_{12}=\mathcal{X}_{1}%
\cup\mathcal{X}_{2}-\mathcal{X}_{1}\cap\mathcal{X}_{2}.
\]
As a trivial example, we consider the [[4, 2, 2]] code. Wherein, $g_{1}%
=Y_{1}Y_{2}Y_{3}Y_{4}$, $g_{2}=Z_{1}Z_{2}Z_{3}Z_{4}$, $Z_{1}=Z_{2}Z_{3}$ and
$Z_{2}=Z_{1}Z_{2}$. It is easy to show that
\[
\mathcal{X}_{12}=\{Z_{1}X_{2}Y_{3}\text{, }Z_{2}X_{3}Y_{4}\text{, }X_{1}%
Z_{3}Y_{4}\text{, }X_{1}Z_{2}Y_{4}\}.
\]
Notably, here $2=d\neq$ $D=3$.

In conclusion, we constructed sets of the undetermined, either pure or mixed,
states. Specifically, we establish the connection between the logic states of
SQECCs and the pure undetermined states through the error correcting power of
the SQECCs. With the undeterminedness property, the reason that the SQECCs are
capable of quantum cryptography tasks becomes trivial. Several applications of
the undetermined states are given in this article. Nevertheless, the
generalized multi-level undetermined states are still open questions for
further studies.

\section*{Acknowledgments}

The author LYH would like to thank Prof. Feng-Li Lin for stimulating
discussion. He also acknowledges support from National Science Council of the
Republic of China under Contract No. NSC.96-2112-M-033-007-MY3. The author MHH
receives support from NSF Grant No.~ECS-0507270. LYH and MHH are also
partially supported by the Physics Division, National Center of Theoertical Sciences.


\begin{thebibliography}{99}                                                                                               %


\bibitem {1}R. Horodecki, P. Horodecki, M. Horodecki, K. Horodecki, e-print
arXiv: quant-ph/0702225.

\bibitem {2}N. Linden, S. Popescu, and W. K. Wootters, Phys. Rev. Lett.
\textbf{89}, 207901 (2002).

\bibitem {21}N. Linden and W. K. Wootters, Phys. Rev. Lett. \textbf{89},
277906 (2002).

\bibitem {22}L. Di\'{o}si, Phys. Rev. A \textbf{70}, 010302 (2004).

\bibitem {4}S. N. Walck and D. W. Lyons, Phys. Rev. Lett. \textbf{100}, 050501 (2008).

\bibitem {5}S. N. Walck and D. W. Lyons, Phys. Rev. A \textbf{79}, 032326 (2009).

\bibitem {NC}M. A. Nielsen and I. L. Chuang, \textit{Quantum Computation and
Quantum Information} (Cambridge University Press, Cambridge, 2000).

\bibitem {qbc}H.-K. Lo and H. F. Chau, Phys. Rev. Lett. \textbf{78}, 3410 (1997).

\bibitem {3}M. Hillery, V. Bu\v{z}ek, and A. Berthiaume, Phys. Rev. A
\textbf{59}, 1829 (1999).

\bibitem {Ins}S.-J. Qin, F. Gao, Q.-Y. Wen, and F.-C. Zhu, Phys. Rev. A
\textbf{76}, 062324 (2007).

\bibitem {513}R. Laflamme, C. Miquel, J. P. Paz, and W. H. Zurek, Phys. Rev.
Lett. \textbf{77}, 198 (1996).

\bibitem {713}A. M. Steane, Phys. Rev. Lett. \textbf{77}, 793 (1996).

\bibitem {Gottesman}D. Gottesman, Phys. Rev. A. \textbf{61}, 042311 (2000).

\bibitem {DG97thesis}D.~Gottesman, Ph.D thesis, Stabilizer codes and quantum
error correction, California Institute of Technology (1997).

\bibitem {pla}M.-Y. Ye and X.-M. Lin, Phys. Lett. A \textbf{372} 4157 (2008).

\bibitem {0602096}M. Hein, W. D\"{u}r, J. Eisert, R. Raussendorf, M. Van den
Nest, and H.-J. Briegel, quant-ph/0602096.

\bibitem {cc}D. Markham, B. C. Sanders, Phy. Rev. A \textbf{78} 042309 (2008).

\bibitem {cjp}L.-Y. Hsu and W.-T. Yen, Chin. J. Phys, \textbf{48} 138 (2010).
\end{thebibliography}

\end{document}